# Observation and control of shock waves in individual nanoplasmas

Daniel D. Hickstein[1], Franklin Dollar[1], Jim A. Gaffney[3], Mark E. Foord[3], George M. Petrov[4], Brett B. Palm[2], K. Ellen Keister[1], Jennifer L. Ellis[1], Chengyuan Ding[1], Stephen B. Libby[3], Jose L. Jimenez[2], Henry C. Kapteyn[1], Margaret M. Murnane[1], and Wei Xiong[1]

[1] Department of Physics, University of Colorado and
JILA, National Institute of Standards and Technology and University of Colorado, Boulder, CO 80309
[2] Department of Chemistry and Biochemistry and CIRES, University of Colorado, Boulder, CO 80309
[3] Physics Division, Physical and Life Sciences, Lawrence Livermore National Laboratory, Livermore, CA 94550
[4] Plasma Physics Division, Naval Research Lab, Washington, DC 20375

ABSTRACT

In a novel experiment that images the momentum distribution of individual, isolated 100-nm-scale plasmas, we make the first experimental observation of shock waves in nanoplasmas. We demonstrate that the introduction of a heating pulse prior to the main laser pulse increases the intensity of the shock wave, producing a strong burst of quasi-monochromatic ions with an energy spread of less than 15%. Numerical hydrodynamic calculations confirm the appearance of accelerating shock waves, and provide a mechanism for the generation and control of these shock waves. This observation of distinct shock waves in dense plasmas enables the control, study, and exploitation of nanoscale shock phenomena with tabletop-scale lasers.

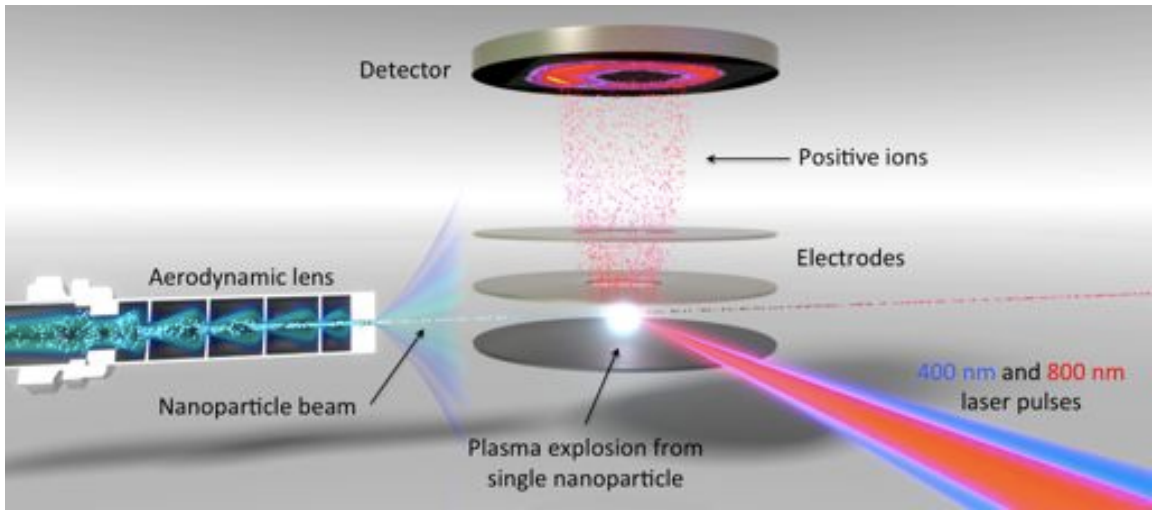

FIG 1. The apparatus for imaging shock waves in individual nanoplasmas. An aerodynamic lens focuses nanoparticles into a high-vacuum chamber where they are irradiated by a series of two time-delayed laser pulses. The first pulse creates an expanding nanoplasma, while the second pulse further heats the plasma, causing a pressure increase, which leads to shock wave formation. The resulting photoion momentum distribution is projected onto a microchannel-plate detector using three electrodes in a velocity-map-imaging geometry [1].

Nanoscale plasmas (nanoplasmas) offer enhanced laser absorption compared to solid or gas targets [2], enabling high-energy physics with table-top-scale lasers. Indeed, previous experimental studies have observed the production of high-energy ions [3] and even nuclear fusion [4] in laser-irradiated nanoplasmas. For more than a decade, theoretical studies have predicted that shock waves can be generated in nanoplasmas, and that these nanoplasma shock waves might allow for the practical generation of quasi-monoenergetic high-energy ions, neutrons from fusion processes, or ultrafast X-ray bursts [5–7].

An analytical study by Kaplan et al. suggests that shocks should be a common phenomenon in expanding nanoplasmas, requiring only a plasma density distribution that is highest in the center and decays smoothly towards the edges [6]. Similarly, Peano et al. [7,8] used numerical simulations to show that the density profile of the precursor nanoplasma would dictate the properties of the shock. In particular, they demonstrated that a weak laser pulse could be used to shape the density profile to so that a second, stronger laser pulse could produce more intense shock waves.

In contrast to the theoretical studies, which model a single nanoplasma, previous experimental studies of nanoplasmas [3,9] used laser focal volumes that contained many particles, thereby simultaneously irradiating nanoparticles of different sizes and with different laser intensities. As we show in this work, the kinetic energy of the shock wave depends on the plasma size and the laser intensity. Thus, studies that probe many nanoparticles simultaneously would create an ensemble of shock waves with different kinetic energies, thereby obscuring their identification as shocks.

In this Letter, by imaging individual laser-irradiated nanoparticles, we remove the size and intensity averaging present in previous studies, which allows us to clearly observe nanoscale plasma shock waves. Furthermore, we demonstrate that these shock waves can be controlled by using a laser pulse to shape the plasma density profile. Finally, we present hydrodynamic simulations that provide a mechanism for the generation and control of shock waves in nanoplasma.

Our observation of shock waves in nanoplasma is enabled by a unique experiment [Fig. 1] that can detect photoions from the nanoplasma generated from a *single* laser-irradiated nanoparticle. Nanocrystals of NaCl, KCl, KI, or $NH_4NO_3$ with diameters of ~100 nm are created using a compressed-gas atomizer and introduced into the vacuum chamber using an aerodynamic lens. A plasma is formed via illumination of a particle with a tightly focused 40-fs laser pulse (wavelength of either 400 nm or 800 nm) with an intensity that is adjusted between $3 \times 10^{13}$ and $4 \times 10^{14}$ W/cm². The angle-resolved energy distribution of the ions created by the expanding nanoplasma is recorded using a velocity-map-imaging (VMI) photoion spectrometer [1,10–12] that records a two-dimensional projection of the photoion angular distribution (PAD).

Because the laser focal spot is small compared to the spacing between the nanoparticles, we probe, on average, one nanoplasma every 40 laser-shots (See Supplemental Material [13] for complete experimental details). In all laser-irradiated nanoparticle experiments, each nanoparticle will experience a different laser intensity depending on where it is located in the laser focus, which leads to intensity averaging effects if each PAD contains ions from many nanoparticles, as was the case in previous nanoplasma studies [14–16]. However, in this experiment, each PAD corresponds to a single particle and, although the intensity cannot be precisely controlled for each particle, no intensity averaging takes place within a single PAD. This allows for the observation of previously undiscovered physical processes, even those that are exquisitely sensitive to laser intensity, particle size, or particle composition.

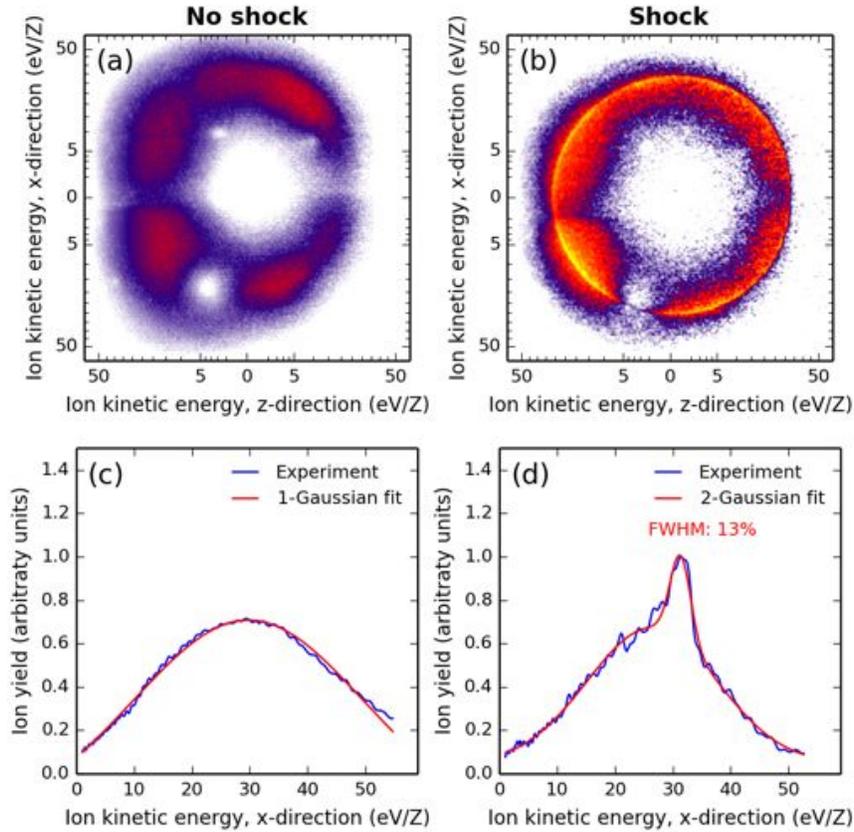

FIG 2. Observation of a shock wave from an individual nanoplasma. (a) The photoion angular distribution (PAD) from a single $NH_4NO_3$ nanoparticle irradiated with a pulse of 400 nm light followed by pulse of 800 nm light typically displays a broad ion distribution. Here the laser propagates in the z-direction (right-to-left) and is linearly polarized in the x-direction. The angular features are due to the inhomogeneous responsivity of the imaging detector. (b) If the particle size, laser intensity, and laser pulse time-delay are tuned appropriately, a sharp shock wave (orange and yellow) appears in addition to the broad ion distribution. (c) The radial energy distribution of the typical nanoplasma explosion can be fit by a single broad Gaussian function. (d) The shock wave manifests as an additional sharp peak, which can be fit by a second Gaussian function with a narrow energy spread.

In our experiment, when the peak laser intensity is below $5 \times 10^{13}$ W/cm$^2$, the PADs contain only 100 or fewer ions, corresponding to the ionization of the residual $N_2$ and $H_2O$ gas that flows with the particles through the aerodynamic lens. However, when the laser intensity is increased above $\sim 5 \times 10^{13}$ W/cm$^2$, we observe some PADs that contain more than 10$^4$ ions, indicating plasma formation in a single nanoparticle [Fig. S1] [13]. Indeed, in this intensity regime, solid nanoparticles are rapidly (<1 ps) converted into dense nanoplasmas through the following mechanism [16–18]. First, the strong laser field causes some of the atoms to ionize through tunnel ionization [9], liberating about one electron per atom within a few tens of fs [19]. These free electrons are accelerated by the strong laser field and then drive further rapid ionization through electron impact ionization [16]. The electrons continue to be driven by the laser field and absorb energy through collisions with the ions [20], reaching high temperatures.

When the laser intensity is increased above $1 \times 10^{14}$ W/cm$^2$, shock waves appear in approximately 10% of the nanoparticle PADs [Fig. 2, S1, and S3]. Each shock wave manifests as a sharp ridge on top of a broad photoion distribution. The ion kinetic energy of each shock ranges from 15 eV/$Z$ to 50 eV/$Z$, where $Z$ is the charge state of the positive ion. However, each individual shock is quasi-monoenergetic, with an energy spread of less than 15% [Fig. 2(b)].

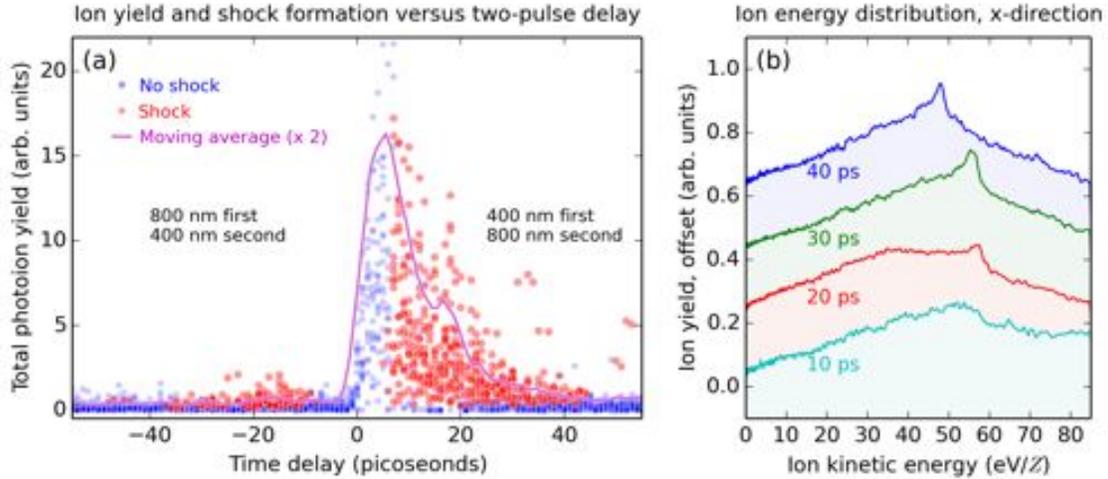

FIG. 3. Control of shock wave formation using two laser pulses. (a) Each dot indicates a single nanoplasma explosion of an individual NH$_4$NO$_3$ nanoparticle as a function of the delay between the 400 nm and 800 nm laser pulses. The first pulse forms a slowly expanding nanoplasma, and the second pulse causes a rapid pressure increase inside of the nanoplasma, which leads to the formation of a shock wave. When the delay between the two pulses is greater than 7 ps, shock waves are formed. The ion yield is higher when the 400 nm pulse precedes the 800 nm pulse because the 800 nm pulse is more effective at heating the expanded nanoplasma. (b) As the relative time delay between the laser pulses is increased, the shocks become more pronounced. For comparison purposes, we display shocks with energies ~50 eV, though the shocks from different nanoparticles range between ~15 and 50 eV.

The formation of the shock waves is sensitive to both the physical size and chemical composition of the particle. We observe that larger nanoparticles are more likely to create shock waves [Fig. S2], which can be explained by the fact that larger nanoplasmas absorb more energy from the laser field [2]. Using a single laser pulse, shocks are observed in a variety of compounds, including KI, NaCl, and KCl, and the threshold laser intensity required to create shocks scales roughly with the ionization potential of the compounds [Fig. S3], as expected for the onset of tunnel ionization [19], and in agreement with the relative ionization yields observed in single-particle mass spectroscopy experiments [21]. For NH$_4$NO$_3$, the compound with the highest ionization potential in this study, no shocks are observed in the single-pulse experiment [Fig. S2], making it the ideal example case for demonstrating the two-pulse shock generation scheme.

Two laser pulses with an appropriate relative time delay can be used to create shock waves in *all* of the nanoparticles investigated in this study, including NH$_4$NO$_3$. The likelihood of shock formation depends critically on the time delay between the first and second pulses. The minimal time delay for shock creation coincides with the peak in the total photoion yield, which occurs around 7 ps [Fig. 3a]. Similarly, the maximum time delay for shock production occurs near 45 ps, corresponding to the end of the enhanced ion yield. Previous studies [15,22] observed a similar dependence of the photoion yield on time delay during the two-pulse irradiation of nanoparticles (although they did not observe shocks) and attributed this behavior to the increased absorption of the second laser pulse caused by the expansion of the plasma following the first laser pulse.

The expansion of our nanoplasma into the vacuum is significantly slower than previous studies due to the large size of the nanoparticles and can be estimated using the ion sound speed [9] $v_{\text{expand}} \approx \sqrt{\frac{ZkT_e}{m_i}}$, where $Z$ is the charge of the ions, $m_i$ is the mass of the ions, and $kT_e$ is the electron temperature of the plasma. For $N^{+1}$ with a temperature of 10 eV, a 100 nm diameter particle would double in size in 6 ps, in good agreement with the 7 ps delay for shock wave formation.

After the nanoparticle is irradiated by the first laser pulse, the resulting plasma expands, and its density assumes a radial profile that decays smoothly into vacuum. Energy absorption peaks when the electron density of the plasma is near the critical density [9,23], the density at which electrons in the plasma are driven resonantly by the laser field. As the plasma expands, the volume of plasma near the critical density expands, enhancing energy absorption. Eventually, the entire nanoplasma drops below the critical density and light absorption is diminished. Thus, the arrival time of the second pulse relative to the first determines the amount of energy absorbed. The similar timescale of ion yield enhancement and shock formation suggests that the two effects share a common mechanism: the expansion of the plasma between the first and the second pulses is crucial for the formation of shocks in the two-pulse experiment.

The time delay between the laser pulses not only determines the presence of shocks, but also determines the fraction of ions that become part of the shock wave. The shocks produced with time delays of ~10 ps involve a small fraction of the ions, while the shock generated using time delays of >15 ps contain a much larger fraction of the total ions [Fig. 3(b) and S4]. This indicates that the first pulse is shaping the plasma density to achieve a density profile that is better optimized for shock wave propagation and generation of quasi-monoenergetic ions. Thus, this demonstrates that it is possible to control shock waves in plasmas by actively sculpting the plasma density profile using a femtosecond laser pulse.

To investigate the mechanism for shock formation we employ numerical hydrodynamic simulations using the radiation hydrodynamics code HYDRA [24]. Details about the simulations are described in the Supplemental Material [13]. We simulate the interaction of two time-delayed laser pulses (each with an intensity of $4.9 \times 10^{14}$ W/cm$^2$) with a 100 nm diameter nanoparticle composed of NaCl. A prominent shock waves is observed for time delays between 5 and 35 ps [Fig. 4 and S5]. In addition, the hydrodynamic simulations accurately reproduce measured ion kinetic energies. The good agreement between simulated and observed ion energies indicates that the hydrodynamic calculations capture of the physics of the plasma expansion.

The hydrodynamic calculations suggest a simple mechanism for shock formation [Fig 4]. After the first pulse expands the cluster, the second laser pulse is absorbed in a relatively thin shell at the critical density [Fig. 4(a)]. The resultant heating produces a large pressure increase in that region that drives material away from the absorption region [Fig. 4(b)]. This shock wave reflects from the center of the plasma, resulting in a population of high-velocity ions at small radius [Fig. 4(b)]. These ions drive an outward-moving shock [Fig. 4(c)], and the associated density increase produces a peak in the ion kinetic energy distribution [Fig. 4(d)]. The energy of the shock is determined both by the energy imparted by the laser and the work required for the high-velocity ions to accelerate the material ahead of them as they move to large radius. Similar to previous theoretical predictions [5,6,25] of shock formation in the Coulomb explosion of small clusters, the shock formation occurs when faster particles towards the interior of the plasma overrun slower particles in the exterior of the cluster. However, in this case, the velocity differential is caused by the preferential absorption of light near the critical density, which creates a ridge of high pressure. In contrast to studies conducted in the Coulomb explosion regime, where the shock is formed on the time scale of 100 fs, the shocks in these hydrodynamic explosions take ~50 ps to form.

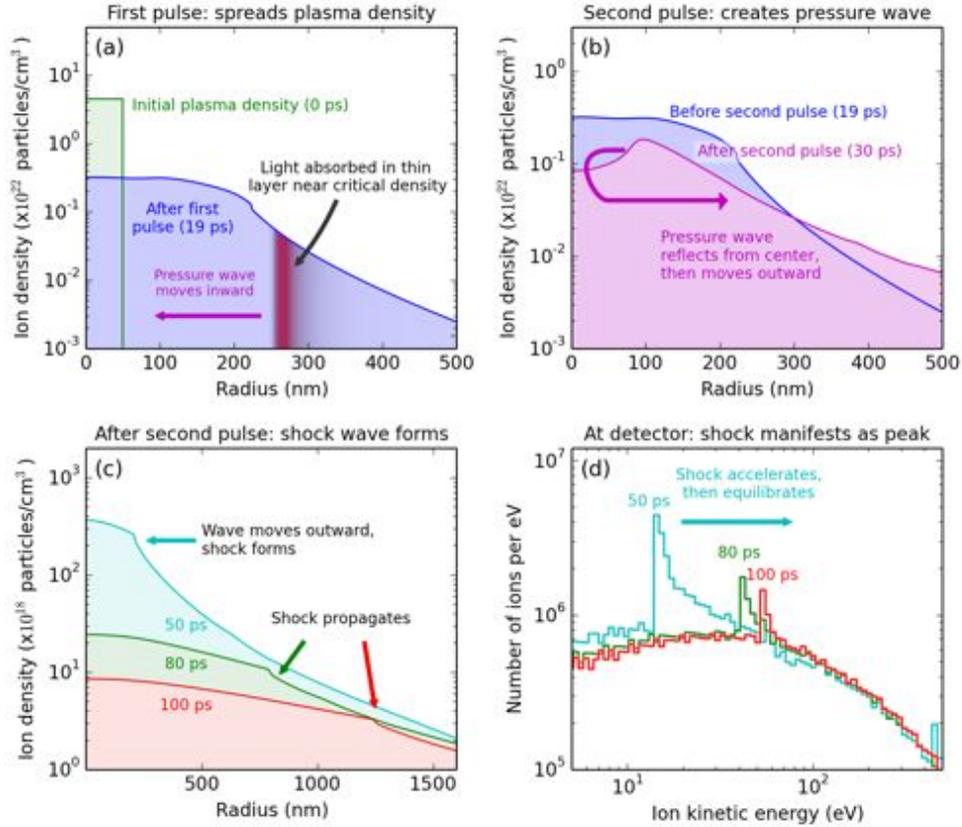

FIG. 4. The mechanism for two-pulse shock wave formation, as revealed by numerical hydrodynamics simulations using HYDRA [24]. (a) The first laser pulse creates a plasma, which expands into the surrounding vacuum. A small single-pulse shock wave can be seen near 230 nm. Energy from the second laser pulse is preferentially absorbed in a layer near the critical density at ~270 nm. (b) The heating from the second laser pulse creates a pressure wave that moves inwards, reflects from the center, and then moves outwards. (c) The pressure wave moves outwards supersonically and accelerates slower material in front of it, creating a shock wave, which is seen as a small step in the ion density distribution. (d) In the kinetic energy distribution of the ions, the shock wave is a sharp peak which is accelerated by the pressure gradient, eventually settling at a final kinetic energy of a few 10s of eV after ~100 ps. It is this asymptotic kinetic energy distribution that is recorded by the spectrometer in the experiment. In this simulation, both laser pulses are modeled as 50 fs, 800 nm pulses with intensities of $4.9 \times 10^{14}$ W/cm$^2$.

Interestingly, the hydrodynamic simulations show a shock wave that accelerates as it moves outwards [Fig. 4(c)], which is most easily seen by the temporal increase in energy of the peak in the ion distribution [Fig. 4(d)]. In our simulations, the shock velocity increases by more than twofold during this acceleration period. The mechanism for such shock acceleration is well known, and stems from the radially decreasing density profile in the background plasma. In the classic Sedov-Taylor Waxman-Shvarts [26] analysis of the problem, acceleration is seen for steep density gradients. Simple dimensional analysis scaling laws [27], which describe the asymptotic behavior of the shock, are in good agreement with the simulated shock acceleration once we account for the fact that our plasma density is rapidly decreasing with time. Accelerating shock waves are of great interest in astrophysics and, consequently, experiments have been proposed to investigate such shocks in the laboratory setting [28]. We believe that study is the first realization of such an experiment and could serve as a versatile platform for studying shocks propagating through customizable density gradients.

Here we presented the first measurements of individual nanoplasmas, demonstrating a new method for studying laser-plasma interactions, which can be implemented using a tabletop apparatus and at a high repetition rate. By characterizing the momentum distribution of individual nanoplasmas, we make the first observation of plasma shock waves on the nanometer scale, confirming a decade of theoretical predictions [5,6,25]. By adjusting the time delay between two laser pulses, the creation and strength of the shock wave was varied in a controllable manner. Furthermore, because these shocks are produced in plasmas with temperatures of just ~10 eV, this experiment potentially enables a compact, inexpensive method for studying a relatively unexplored regime of low-temperature nanoplasmas.

D.D.H., W.X., F.D., C.D., K.E.K., J.L.E., M.M.M., and H.C.K. acknowledge support from the DOE Office of Fusion Energy Sciences DE-SC0008803. B.B.P. and J.L.J. thank DOE DE-SC0006035 and NOAA NA13OAR4310063 for support. J.A.G. would like to thank Marty Marinak (LLNL) for his assistance with HYDRA simulations. S.B.L., M.E.F, and J.A.G. acknowledge support from the DOE Office of Fusion Energy, HED Laboratory Plasmas program under grant AT5015033. Lawrence Livermore National Laboratory is operated by Lawrence Livermore National Security, LLC, for the U.S. Department of Energy, National Nuclear Security Administration under Contract DE-AC52-07NA27344. G.M.P. acknowledges support by the Naval Research Laboratory 6.1 Base Program.

# Observation and control of shock waves in individual nanoplasmas

Daniel D. Hickstein[1], Franklin Dollar[1], Jim A. Gaffney[3], Mark E. Foord[3], George M. Petrov[4], Brett B. Palm[2], K. Ellen Keister[1], Jennifer L. Ellis[1], Chengyuan Ding[1], Stephen B. Libby[3], Jose L. Jimenez[2], Henry C. Kapteyn[1], and Margaret M. Murnane[1], Wei Xiong[1]

[1]Department of Physics, University of Colorado and
JILA, National Institute of Standards and Technology and University of Colorado, Boulder, CO 80309
[2]Department of Chemistry and Biochemistry and CIRES, University of Colorado, Boulder, CO 80309
[3]Physics Division, Physical and Life Sciences, Lawrence Livermore National Laboratory, Livermore, CA 94550
[4]Plasma Physics Division, Naval Research Lab, Washington, DC 20375

## Experimental Details

The experimental apparatus consists of a nanoparticle aerosol source coupled to a velocity-map-imaging (VMI) photoion spectrometer [1–3]. Starting with an aqueous solution, a compressed-gas atomizer (TSI Inc. model 3076) generates an aerosol consisting of droplets with an average diameter of approximately 1 um. The water in the droplets then evaporates, leaving behind nanocrystals with a diameter that depends on the sample concentration. The nanocrystals are typically 100 nm in diameter, and have approximately spherical shape [4]. Samples of NaCl, KI, KCl, and $NH_4NO_3$ were obtained from Fisher Scientific and diluted in ultrapure (>18 MΩ resistivity) water. The samples were diluted to the same volume concentration (0.12%) to assure that the aerosol particles of different compositions would be the same size. The nanocrystals are focused using an aerodynamic lens (Aerodyne Research, Inc.), which uses a series of ~2 mm apertures to collimate the nanocrystal aerosol into a ~0.5 mm beam.

The collimated aerosol beam passes through a 1.5 mm diameter skimmer and into the differentially pumped photoionization chamber, which reaches a pressure of $1 \times 10^{-6}$ mbar (base pressure $5 \times 10^{-10}$ mbar). The nanoparticles are then ionized by an intense (~$10^{14}$ W/cm$^2$) 800 nm, 40 fs laser pulse, derived from a Ti:sapphire regenerative amplifier (KMLabs) operating at 1 kHz. The pre-pulse contrast is greater than 250:1 as measured with a photodiode, and the pulse energies are typically 5 to 100 $\mu$J.

The 10-mm laser beam is focused with a 30 cm lens to reach an estimated FWHM focal spot diameter of 25 um. The resulting volume of the interaction region (assuming a 0.5 mm collimated aerosol beam) is $2.5 \times 10^{-7}$ cm$^3$. With an estimated aerosol density of $1 \times 10^5$ particles/cm$^3$, the laser pulse will interact with a particle every ~40 laser pulses, for an average of 25 hits per second at a 1 kHz repetition rate. Thus, even with millisecond exposure times, we can identify the photofragments originating from a single nanocrystal.

The actual hit rate observed experimentally depends on the volume of the laser focus that is above the threshold intensity for plasma formation and thus depends on the power of the incident laser beam.

For the two-pulse experiment, we used a BBO crystal to generate ~40 fs pulses of 400 nm light. The 400 nm and 800 nm pulses were delayed in time using a computer-controlled delay stage in a Mach-Zehnder configuration. The photoions are accelerated towards the microchannel-plate phosphor detector using three electrodes in a velocity-map-imaging geometry [1], and the photoion distribution is recorded using a CCD camera (Allied Vision Technologies).

## Analysis of shock formation with particle size

For the size-selected experiments, the nanoparticles were passed through a differential mobility analyser (TSI Inc., models 3081 and 3085). To quantify the relative number of particles created by the atomizer, we used a scanning-mobility particle-sizer (SMPS) spectrometer consisting of the differential mobility analyser connected to a condensation particle counter (TSI Inc., model 3775).

## Hydrodynamic simulations using HYDRA

To investigate the mechanisms for shock wave formation, hydrodynamic simulations were performed using version 9.0 of the HYDRA radiation-hydrodynamics code [6] for NaCl nanoparticles of 100 nm radius. The HYDRA calculations were performed in 1D, by assuming spherical symmetry, using the 3D version of HYDRA. We note that this version of HYDRA, which uses finite elements methods (rather than finite difference methods) for transport calculations, is more appropriate for these plasmas. Simulations used an adaptive one (radial) dimensional Lagrangian mesh, which uses a smaller grid zones near regions of rapidly changing plasma density. Thus, the calculations give a reliable description of sharp spatial features (such as shock waves). Furthermore, the calculations include models for laser energy deposition via inverse bremsstrahlung (IB) along with accurate equation of state and thermal transport quantities.

The hydrodynamic calculations confirm that a shock wave is produced in the expanding nanoplasma and that this shock manifests as a step in the plasma density as well as a spike in the ion kinetic energy distribution [Fig. 4, S5, and S6]. The simulations indicate that the simulated spike in the ion kinetic energy distribution is quite sensitive to the parameters of the laser pulse; in this work we have chosen intensities to match the experimentally observed ion energy of ~20 eV with the 20 ps delay between the two laser pulses. This results in a simulated laser intensity of $5 \times 10^{14}$ W/cm$^2$, which is consistent with the estimated experimental laser intensity.

In HYDRA, the initial ionization of cold material is assumed to occur instantaneously. Using the ADK ionization model [7] we calculate that for our laser parameters, initial ionization of the particle occurs well before the peak of the laser pulse,[2] after which IB takes over and the HYDRA will model the plasma adequately. The small size of the nanoparticles combined with the relatively high temperatures created by the second laser pulse mean that the thermal electron mean free path is larger than the spatial mesh zone size, making non-local energy transport important. A model for non-local energy transport is implemented in HYDRA [8], and we include this in the presented simulations.

The HYDRA calculations make the additional approximations 1) that the net charge of the plasma is small and 2) that the electron energy distribution can be described by a Maxwellian distribution, both of which are valid for our nanoparticles. Regarding the first approximation, we show below (in the section titled "Plasma quasi-neutrality") that, due to the low laser intensities used in this study, only a small fraction of the electrons will be able to leave the plasma during the laser pulse. Therefore, the plasma is nearly net-neutral and can be modeled as a hydrodynamic expansion. The second approximation, that the electrons energy distribution can be described as Maxwellian, is valid because the excited electrons will thermalize on a very rapid timescale compared to the picosecond timescale of the HYDRA simulations. The electron thermalization time can be estimated from the electron-electron energy exchange rate [9]

$$\tau_e = 3.44 \times 10^5 \frac{T_e^{3/2}}{N_e \lambda} \text{ sec,}$$

where $T_e$ is the electron temperature, $N_e$ is the density of the electrons, and $\lambda \approx 3$ is the Coulomb Logarithm. Electron thermalization will be the slowest at the time of the second pulse, when electron temperatures are high and the plasma is lower density than during the first pulse. In the region of high-pressure that drives the observed shock, typical simulated conditions are $T_e = 20$ eV and $N_e = 5 \times 10^{21}$ cm$^{-3}$. The timescale for electron thermalization is then ~0.6 fs, very fast compared to the hydrodynamic motion of interest in this study.

Consistency between the simulated velocity distribution and the experimentally recorded velocity distribution has been checked by numerically propagating the velocity distribution towards a "detector" using a simulated electric field. This analysis used the simulated velocity profile taken 80 ps after the second laser pulse, the point at which we estimate the VMI field separates electron and ion clouds, allowing the ions to propagate to the detector without further interactions. The true process of image formation is complex, and has not yet been modeled, however we have confirmed that sharp structures in VMI images correspond to peaks in the kinetic energy distribution.

## Supplemental Figures

The following supplemental figures present the observation of shock waves from a single laser pulse in NaCl (Fig. S1), the dependence of shock wave formation on particle size (Fig. S2), the increase of shock wave formation with laser intensity (Fig. S3), the effect of time-delay on the ion yield of the shock wave (Fig. S4), the simulated dependence of the shock wave on time-delay (Fig. S5), and the simulated dependence of the shock wave on the intensity of the second laser pulse (Fig. S6).

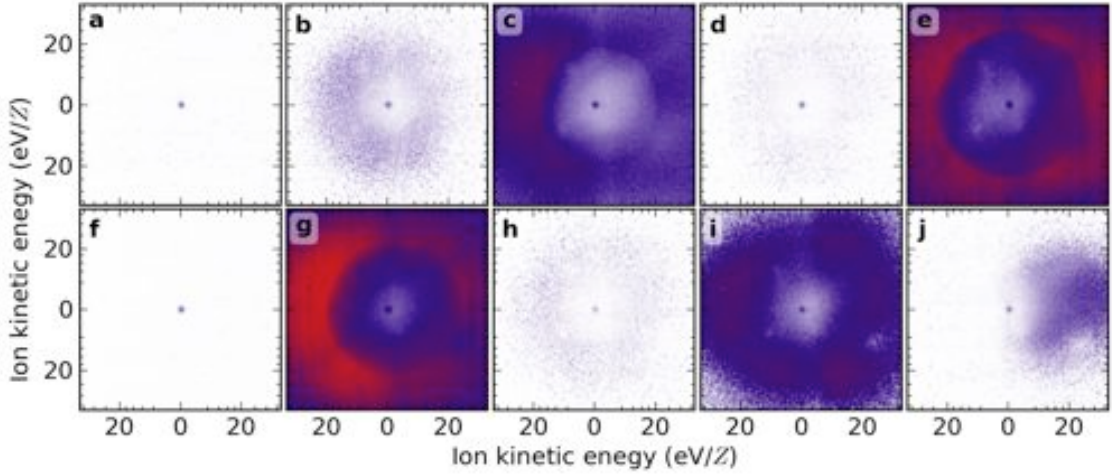

FIG. S1. Photoion angular distributions from nanoplasma explosions of NaCl nanoparticles with a single 800 nm laser pulse. Panels a and f show no particles, while panels b, d, h, i, and j show diffuse plasma explosions. Frames c, e, and g display shock waves that were generated from particles subjected to laser intensities sufficient to initiate shock formation. The center of the frame corresponds to ions with zero kinetic energy, and the laser polarization is in the vertical direction. The laser propagates from left to right.

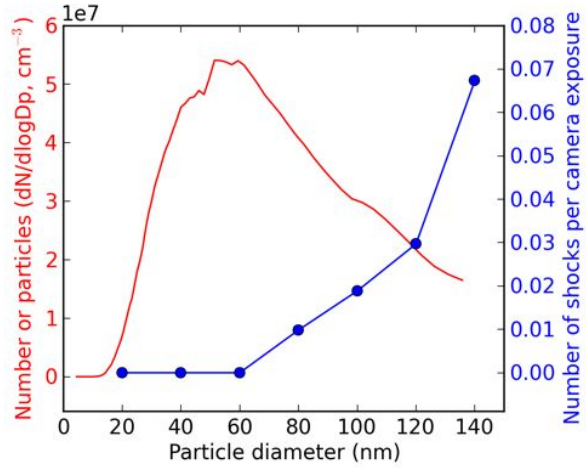

FIG. S2. Effect of particle diameter of the formation of shock waves. With a single 800 nm laser pulse with an intensity of $3 \times 10^{14}$ W/cm², the frequency of creating shocks in NaCl particles increases as the particle diameter becomes larger, even though the concentration of particles is decreasing for particle sizes larger than ∼60 nm. Larger nanoparticles absorb more energy per atom and can therefore create shocks throughout a larger region of the laser focal volume.

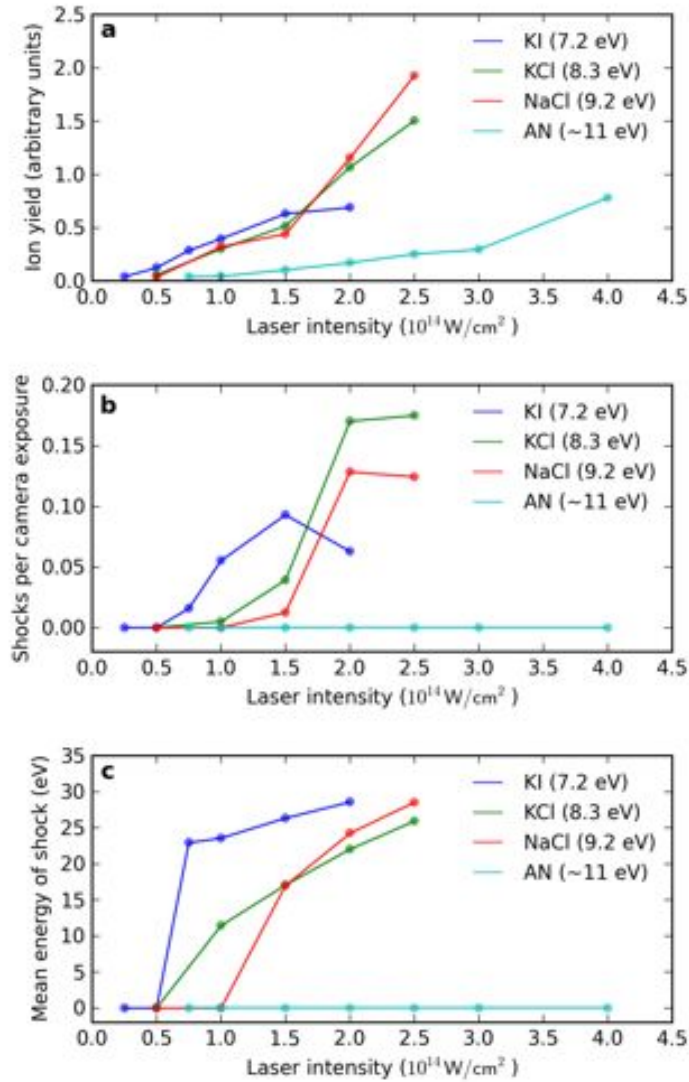

FIG. S3. Ion yield and shock wave formation versus laser intensity. (a) The ion yield suddenly increases with laser intensity, which is typical for avalanche ionization, a process that ionizes most of the atoms in the nanoparticle and forms a dense plasma [10]. The threshold laser intensity increases with the ionization potential of the species (shown in parenthesis). The ionization potentials were obtained from the NIST webbook [11], except for $NH_4NO_3$ (ammonium nitrate, AN), which does not have a documented ionization potential, but is expected to have one similar to other compounds featuring a nitrate ($NO_3$) moiety, which have ionization potentials between 11 and 12 eV. (b) Ammonium nitrate requires higher laser intensity to achieve comparable ion yields and does not show any shock rings in this one-pulse experiment. The number of shocks generated per camera exposure increases rapidly at a threshold intensity, quickly saturating. (c) The average energy of the shock rings increases with laser intensity until reaching an asymptotic value of ~30 eV/$Z$.

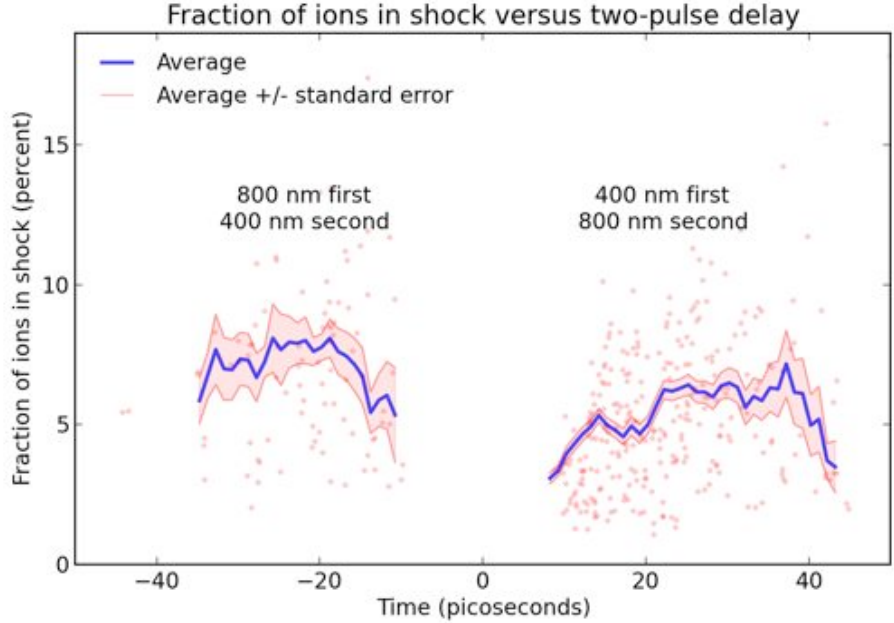

FIG. S4. The relative fraction of ions in the shock wave controlled with two-pulse delay. Each dot corresponds to a single shock wave generated from a $NH_4NO_3$ nanoparticle. The relative number of ions in each shock wave compared to the total number of ions is estimated by fitting the ion distribution with two Gaussian functions [Fig. 2(d)]: one wide Gaussian for the broad "background" ions and a narrow Gaussian for the ions in the shock wave. At time delays of 10 ps, the shock waves do not contain many ions compared to the background. However, the fraction of ions in the shock is enhanced at an optimal time delay around 30 ps.

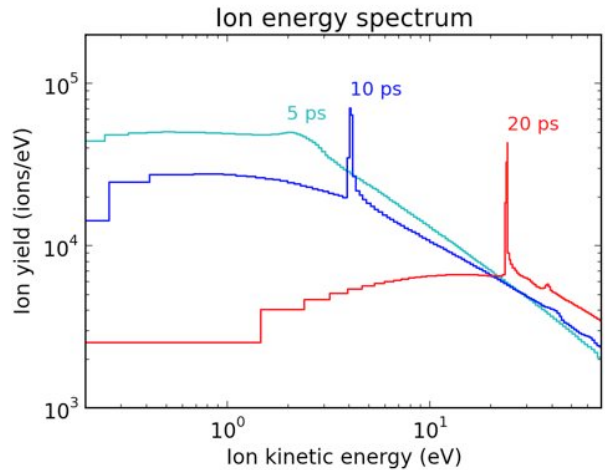

FIG. S5. Numerical simulations of two-pulse accelerating shock production in NaCl. Hydrodynamic simulations completed using the HYDRA [6] radiation-hydrodynamics code reveal that two time-delayed 800 nm laser pulses, both with intensities of $4.9 \times 10^{14}$ W/cm², will generate quasi-monoenergetic shocks in the ion kinetic-energy spectrum only when the time delay is less than ~35 ps. The shock does not show up as a sharp peak for the time delay of 5 ps, but when the time delay is increased to 10 ps, the shock displays as a sharp peak, in agreement with experimental findings

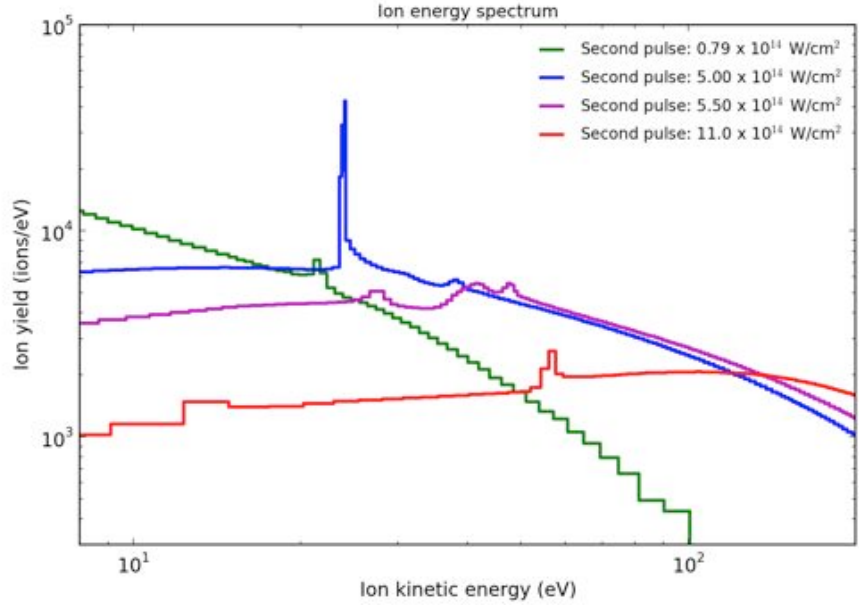

FIG. S6. Simulated scaling of the shock with intensity of the second laser pulse in NaCl nanoparticles. The ion kinetic energy distribution from two laser pulses delayed by 20 ps as calculated by hydrodynamic simulations using the HYDRA software [6]. The intensity of the first pulse is $4.9 \times 10^{14}$ W/cm². Increasing the intensity of the second pulse from $0.79 \times 10^{13}$ to $1.1 \times 10^{15}$ W/cm² increases the energy of the quasi-monoenergetic ions produced by the shock wave by about 5 times. However, the shape of the peak in the ion distribution is extremely sensitive to the laser intensity, with ~10% changes in the laser intensity transforming a single sharp peak into three broad peaks.

## Plasma quasi-neutrality

In contrast to most previous studies of nanoplasmas [10,12–16], this study uses ~100 nm nanoparticles (containing on the order of $10^7$ atoms) versus <10 nm noble gas clusters (containing on the order of $10^3$ or $10^4$ atoms). One important effect of this size difference is that the most of the electrons in these larger plasmas cannot escape the plasma, producing a nanoplasma that has a small charge imbalance (a quasi-neutral plasma). There are several reasons for this difference. First, the excursion distance for laser-field-driven electrons is comparatively smaller for larger nanoplasmas – much less than the cluster radius – preventing the majority of electrons from being driven outside of the ion cloud by the laser field. Second, even if a very small fraction of the electrons leave the plasma, a massive charge builds up, preventing the majority of the electrons from escaping the nanoplasma.

Here we estimate the fraction of electrons that can leave the nanoplasma $\Delta N_e/N_e$, based on energy considerations. When free electrons are driven out of the cluster by the laser electric field, the charge imbalance creates a potential well. If too many electrons leave the cluster, the potential energy $E_{\text{pot}}$ due to excess charge will become larger than the kinetic energy of the escaping electrons $E_{\text{kin}}$, which will prevent further electron escape. One can estimate $\Delta N_e/N_e$ by equating $E_{\text{pot}}$ to $E_{\text{kin}}$, assuming that the excess charge is uniformly distributed within the cluster. Since the potential energy is given by

$$E_{\text{pot}} = \frac{Ne^2}{4\pi\varepsilon_0 R} \frac{\Delta N_e}{N_e}$$

and taking

$$E_{\text{kin}} = 4\frac{e^2}{4\pi\varepsilon_0 a_{\text{Bohr}}} = 100 \text{ eV}$$

we get

$$\frac{\Delta N_e}{N_e} = \frac{4R}{Na_{\text{Bohr}}} = \frac{4R[\text{a.u.}]}{N} = 10^{-3}$$

which corresponds to 0.1% of the electrons leaving the cluster. It should be noted that the small kinetic energy of the electrons (~100 eV, due to the low laser intensity) plays an important role for keeping the electrons inside the cluster. The electrons are simply not energetic enough to overcome the potential barrier, which arises due to charge imbalance.

The predicted 0.1% ionization for large clusters is in sharp contrast to small noble gas clusters used in previous studies. As an example, for $R = 5$ nm $= 94$ a.u., $N = 10^4$, and $E_{\text{kin}} \cong 1 \, KeV$, we get

$$\frac{\Delta N_e}{N_e} \cong \frac{40R}{Na_{\text{Bohr}}} \cong \frac{40 \times 94}{10^4} \cong 0.4,$$

i.e. ~40 % of the electrons leave the cluster. This analysis reveals why small clusters become highly charged and undergo fast Coulomb explosion while larger clusters remain quasi-neutral and undergo a slow hydrodynamic expansion.